\documentclass[a4paper,english,12pt]{article}
\usepackage{jheppub} 
\usepackage{float}
\usepackage{amsmath}
\usepackage{array,arydshln}
\makeatletter

\numberwithin{equation}{section}

\usepackage[T1]{fontenc} 

\usepackage{amssymb}
\usepackage{amsmath}
\usepackage{epstopdf}
\usepackage{dsfont}
\usepackage{multicol}
\usepackage{braket}		
\usepackage{simplewick}
\usepackage{slashed}
\usepackage{dsfont}
\usepackage{titlesec}
\usepackage[title,page]{appendix}
\usepackage[numbers]{natbib}

\numberwithin{equation}{section}

\newcommand{\nn}{\tilde N}
\newcommand{\tr}{\text{Tr}}
\newcommand{\id}{\mathds 1}
\allowdisplaybreaks

\newcommand{\sign}[1]{\text{sign}\left(#1 \right)}

\newcommand{\fm}{F_m}

\newcommand{\te}[1]{\text{#1}}
\renewcommand{\b}{\bold}

\newcommand{\p}{\phi}

\makeatother

\title{IR Dualities in General 3d Supersymmetric ${SU(N)}$ QCD Theories}

\preprint{WIS/09/14-NOV-DPPA}

\author{
Ofer Aharony and Daniel Fleischer
\\
\it{ Department of Particle Physics and Astrophysics,\\
Weizmann Institute of Science, Rehovot 7610001, Israel}
}
\emailAdd{Ofer.Aharony@weizmann.ac.il}
\emailAdd{Daniel.Fleischer@weizmann.ac.il} 

\abstract{
In the last twenty years, low-energy (IR) dualities have been found for many pairs of supersymmetric gauge theories with four supercharges, both in four space-time dimensions and in three space-time dimensions. In particular, duals have been found for 3d ${\cal N}=2$ supersymmetric QCD theories with gauge group $U(N)$, with $F$ chiral multiplets in the fundamental representation, with ${\tilde F}$ chiral multiplets in the anti-fundamental representation, and with Chern-Simons level $k$, for all values of $N, F, {\tilde F}$ and $k$ for which the theory preserves supersymmetry. For $SU(N)$ theories the duals have been found in some cases, such as $F={\tilde F}$ and ${\tilde F}=0$, but not in the general case. In this note we find the IR dual for $SU(N)$ SQCD theories with general values of $N, F, {\tilde F}$ and $k \neq 0$ which preserve supersymmetry.
}

\begin{document}
\maketitle

\section{Introduction}

Twenty years ago today, Seiberg discovered \cite{Seiberg:1994pq} that certain pairs of four dimensional ${\cal N}=1$ $SU(N)$ gauge theories exhibit an IR duality, namely they flow to the same theory at low energies (and, in particular, to the same superconformal field theory at the origin of moduli space). In the following years this was generalized to many other examples of gauge theories with four supercharges in four, three and two space-time dimensions. In three dimensions dualities were first found for $U(N)$ gauge groups rather than $SU(N)$ groups. These dualities were first discovered for theories with $F$ chiral multiplets in the fundamental representation and ${\tilde F}=F$  in the anti-fundamental representation \cite{Aharony:1997gp}. This was then generalized to the case where there is also a Chern-Simons (CS) level $k$ \cite{Giveon:2008zn}, and then to arbitrary values of $N, F, {\tilde F}$ and $k$ for which the theory preserves supersymmetry \cite{Benini:2011mf}, subject to the ${\mathbb Z}_2$-anomaly constraint \cite{Niemi:1983rq,Redlich:1983kn,Redlich:1983dv}
\begin{equation}
\label{quantization}
k+(F-{\tilde F})/2 \in {\mathbb Z}. 
\end{equation}
All of these dualities do not have any rigorous derivation so far, but they pass many consistency checks, and they are all related to each other by various flows.

The 3d dualities for $SU(N)$ gauge theories were only discovered relatively recently  \cite{Aharony:2013dha,Park:2013wta}, and were studied so far only in some special cases such as the non-chiral case $F={\tilde F}$ or the purely chiral case ${\tilde F}=0$. Our goal in this note is to derive dualties for general values of $N, F, {\tilde F}$ and $k \neq 0$, subject to \eqref{quantization}. The method we will use is to start from the known duality with $F={\tilde F}$, and to add real masses for some of the flavors in the fundamental representation to reach the theories with $F \neq {\tilde F}$. By following the same flow in the dual theory, we will find the desired duals.

Note that for $F = {\tilde F}$ one can derive the 3d $SU(N)$ duality by compactifying the corresponding 4d duality on a circle and carefully flowing to low energies \cite{Aharony:2013dha}, but this method is not available for $F \neq {\tilde F}$ since the corresponding 4d theories would be anomalous.

In this note we do not discuss the duals for $SU(N)$ theories with $F \neq \tilde{F}$ and $k=0$. Presumably they can be found by similar methods, or by flowing from the finite $k$ dualities that we describe here, along the lines of \cite{Intriligator:2013lca,Khan:2013aa,Amariti:2014aa}.

We begin in section \ref{theoryA} by describing the flow in theory A, and then in section \ref{theoryB} we describe the same flow in the dual theory B, leading to our desired dual. Finally, in section \ref{flat} we perform a simple test of the duality by comparing the baryonic flat directions on both sides. 

This note is based on \cite{Fleischer:2014aa}, which contains additional details.

\section{The flow in theory A}
\label{theoryA}

We will not review here the background material on 3d ${\cal N}=2$ gauge theories and their known dualities; all the necessary background may be found in \cite{Aharony:1997ab,Intriligator:2013lca,Aharony:2013dha} and references therein.

We begin with the $SU(N)$ duality for $F={\tilde F}$ \cite{Aharony:2013dha}. Theory A is an $SU(N)_{k}$ Yang-Mills-Chern-Simons theory with $F$ chiral superfields $Q^i$ in the fundamental representation $\b F$ and $\tilde{F}=F$ chiral superfields $\tilde Q_{\tilde i}$ in the anti-fundamental representation $\b {\overline F}$. For simplicity, from now on we take $k>0$; results for $k<0$ can be obtained by a  parity transformation. Theory A is dual to theory B which is an $SU(F-N+k)_{-k}\times U(1)_{F-N}$ gauge theory, with $F$ fundamental chiral superfields $q_i$ and $\tilde{F}=F$ anti-fundamental chiral superfields $\tilde{q}^{\tilde i}$, and with $F\times F$ singlets $M^i_{\tilde i}$, coupled by the superpotential $W=M^i_{\tilde i} q_i \tilde{q}^{\tilde i}$. $M^i_{\tilde i}$ map to $Q^i {\tilde Q}_{\tilde i}$ in theory A. The duality can be written in short as:\footnote{We ignore here the global structure of the gauge group on the right-hand side, which is that of $U({\tilde N})$. This is important for the consistency of the duality but will not play any role in this paper.}
\begin{equation}
\begin{aligned}
SU(N)_{k}\quad\longleftrightarrow\quad SU(\nn)_{-k}\times U(1)_{\nn-k}.
\end{aligned}
\end{equation}
where $\nn=F-N+k$. Theory A has a supersymmetric vacuum only for $F-N+k\ge 0$.

The symmetries of theory A are summarized in the following table:
\begin{table}[H]
	\label{tab:tablename}
	\begin{center}
		\begin{tabular}{c|c|cccc}
		\hline

		\hline
		 & $SU(N)$ & $SU(F)_{L}$ & $SU(F)_{R}$ & $U(1)_{A}$ & $U(1)_{B}$ \\
		\hline
			$Q$			& $\b{N}$	&	$\b{F}$&	$\b{1}$&$1$	&$1$		\\
			$\tilde Q$	& $\overline{\b {N}}$	&$\b{1}$	&$\overline{\b{F}}$	&$1$	&	$-1$\\
		\hline

		\hline
		\end{tabular}
	\end{center}
\end{table}

We want to turn on a real mass for $\fm$ chiral flavors, in order to flow to a theory with general values of $F \neq {\tilde F}$. Without loss of generality, we give mass to fundamental flavors, and then integrate them out. Equivalent results for mass flows involving anti-fundamental flavors can be achieved via a charge conjugation symmetry transformation. 

Turning on real masses is equivalent to turning on background scalars in vector superfields corresponding to the global symmetry currents of the global symmetry group $SU(F)_{L}\times SU(F)_{R}\times U(1)_{B}\times U(1)_{A}$. For simplicity, we turn on an equal mass for $F_m$ flavors in the fundamental representation. The mass matrices then have the form:
\begin{equation}
\begin{aligned}
\label{massA}
\hat{m}_Q & =
\begin{pmatrix}
 m\id_{F_m}\\
 &0\cdot \id_{F-\fm}
\end{pmatrix},\\
\hat{{m}}_{\tilde Q} & =0.
\end{aligned}
\end{equation}
The mass deformation breaks $SU(F)_L \to SU(F-F_m) \times SU(F_m) \times U(1)$.
In order to translate this to theory B it is useful to write it in terms of the background global symmetries : using
\begin{align}
\hat m_Q=m_{SU(F)_L}+m_{U(1)_{A}}\cdot \id_{F}+m_{U(1)_{B}}\cdot \id_{F},\\
\hat m_{\tilde Q}=-m^{\dagger}_{SU(F)_R}+m_{U(1)_{A}}\cdot \id_{F}-m_{U(1)_{B}}\cdot \id_{F},
\end{align}
we can write \eqref{massA} as:
\begin{align}
m_{U(1)_{A}}&=m_{U(1)_{B}}=\frac \fm{2F}\cdot m,\\
m_{SU(F)_L} &= \frac \fm F \begin{pmatrix}
	\frac{F-\fm}{\fm}\id_{\fm}\\
	&-\id_{F-\fm}
\end{pmatrix}\cdot m,\\
m_{SU(F)_R} &= 0,
\end{align}
with $\tr(m_{SU(F)_L})=0$ as required.

Classically it is clear that the D-term equations have a solution at $Q={\tilde Q}=0$, with a vanishing scalar in the vector multiplet. In this vacuum $\fm$ chiral superfields in the fundamental representation are massive and may be integrated out. We expect that for many values of $N, F, \fm$ and $k$ this supersymmetric vacuum will survive also in the quantum theory; as usual in IR dualities this is expected to happen whenever the rank of the dual group is non-negative (while otherwise we expect no supersymmetric vacuum). At low energies, in this vacuum we obtain an $SU(N)_{k+\Delta k}$ Chern-Simons theory, coupled to $F-\fm$ fundamentals and $F$ anti-fundamental
chiral superfields, with a shifted CS level:
\begin{equation}
\begin{aligned}
\Delta k=\frac{\fm\,\sign m}{2}.
\end{aligned}
\end{equation}
Note that the low-energy theory still satisfies \eqref{quantization}, as required for consistency. This vacuum preserves the full global symmetry preserved by the mass deformation, but only $SU(F-F_m)\times SU(F)_R\times U(1)_B \times U(1)_A$ acts non-trivially on the low-energy theory (where the $U(1)$'s are some linear combinations of the original $U(1)$'s and the one coming from $SU(F)_L$).

Theory A also has various other flat directions, and in particular flat directions that involve taking some eigenvalues of the scalar $\hat\phi$ in the $SU(N)$ vector multiplet to scale with $m$. These vacua are infinitely far away in the $m\to \infty$ limit and we will not discuss them here; presumably they can also be matched to theory B, as in the $U(N)$ analysis of \cite{Shamir:2010aa}.

\section{The flow in theory B}
\label{theoryB}

Theory B before we turn on the mass deformation is an $SU(F-N+k)_{-k}\times U(1)_{F-N}$ gauge theory with $F,\,\tilde F=F$ fundamental and anti fundamental fields $q,\tilde{q}$ (charged under both $SU(F-N+k)$ and $U(1)$) and $F\times F$ singlets $M$, coupled by the superpotential $W=Mq\tilde{q}$.
We denote $\tilde{N}=F-N+k$, and as previously mentioned we take $k>0$.

The quantum numbers of these fields are as follows: 
\begin{table}[H]
	\label{tab:tablename}
	\begin{center}
		\begin{tabular}{c|c|cccc}
		\hline

		\hline
		 & $SU({\tilde N})\times U(1)$ & $SU(F)_{L}$ & $SU(F)_{R}$ & $U(1)_{A}$ & $U(1)_{B}$ \\
		\hline
			$q$			& $\b{{\tilde N}}_{1}$	&	$\overline{\b{F}}$&	$\b{1}$&$-1$	&$\frac{N}{F-N}$		\\
			$\tilde q$	& $\overline{\b {\tilde N}}_{-1}$	&$\b{1}$	&${\b{F}}$	&$-1$	&	$-\frac{N}{F-N}$\\
			$M$	& $\b 1$	&$\b{F}$	&$\overline{\b{F}}$	&$2$	&	$0$\\
		\hline

		\hline
		\end{tabular}
	\end{center}
\end{table}
\noindent
where the $U(1)_{B}$ charges are chosen as in \cite{Aharony:2013dha} such that the monopoles do not carry this charge (in general one can mix this symmetry with $U(1)_{gauge}$, under which the monopoles are charged due to the CS coupling).

To analyze the corresponding flow to the one we did in theory A, we need to transform the mass matrices to theory B. Using the fact that now
\begin{align}
\hat m_q=-m^\dagger_{SU(F)_L}-m_{U(1)_{A}}\cdot \id_{F}+\frac N{F-N}\,m_{U(1)_{B}}\cdot \id_{F},\\
\hat m_{\tilde q}=m_{SU(F)_R}-m_{U(1)_{A}}\cdot \id_{F}-\frac N{F-N}\,m_{U(1)_{B}}\cdot \id_{F},
\end{align}
we find
\begin{align}
\hat m_q&= m\begin{pmatrix}
\left[\frac {\fm}{2\left(F-N \right)}-1 \right]\id_{\fm}\\
&\frac {\fm}{2\left(F-N \right)}\id_{F-\fm}
\end{pmatrix},\\
\hat m_{\tilde q}&=-\frac {m\,\fm}{2\left(F-N \right)}\id_{F}.
\end{align}
The mesons $M^i_{\tilde i}$ with $i=1,\cdots,F_m$ also acquire a real mass.

We would like to find a vacuum of this mass-deformed theory which matches the one we discussed in theory A -- the global symmetry should not be broken, the mesons $M^i_{\tilde i}$ should have vanishing expectation value, and the global symmetry $SU(F-F_m) \times SU(F)_R\times U(1)_B\times U(1)_A$ should act non-trivially on the low-energy theory. Since the fields that $SU(F-F_m)$ and $SU(F)_R$ act on have a real mass in theory B, it is clear that we only get light charged particles if this is canceled by a vacuum expectation value for some of the eigenvalues of the scalar field $\phi$ in the $SU({\tilde N})\times U(1)$ vector multiplet, that should be equal to $-m {F_m \over {2(F-N)}}$.
So, we want to try to find consistent supersymmetric vacua with this property. We will do this by solving the classical equations of motion,
assuming that in the range of parameters where the theory does not break supersymmetry every classical solution corresponds to a quantum vacuum as well.

We begin with the simplest ansatz in which $\phi = -m {F_m \over {2(F-N)}} \id_{\tilde N}$, such that the quarks charged under $SU(F-F_m)\times SU(F)_R$ all remain massless, while the other $F_m$ quarks are massive. The D-term equation coming from the $U(1)$ factor, including the contributions from the CS term and from the Fayet-Iliopoulos (FI) term induced by integrating out the massive quarks, is (we use the conventions of \cite{wess1992supersymmetry,Intriligator:2013lca})
\begin{equation}
-\left(F-N \right)\frac {m\,\fm}{4\pi\left(F-N \right)}+\frac{\fm\,m\,\sign m}{4\pi}-\sum_{c=1}^{\tilde N} \left[ \sum_{f=F_m+1}^{F} q^{c}_{f}q^{*f}_{c}- \sum_{f=1}^{F} \tilde q_{c}^{f}\tilde q_{f}^{*c} \right]=0.
\end{equation}
Clearly this has a solution with unbroken flavor symmetry for the light quarks if and only if $m > 0$. In this case
the shift in the CS level of the low-energy theory from integrating out the massive quarks is $\Delta k=-\frac \fm 2\sign m=-\frac \fm 2$, so we get at low energies an $SU(\nn)_{-k-\frac \fm 2}\times U(1)_{F-N-\frac \fm 2}$ gauge theory with $F-\fm$ fundamentals and $F$ anti fundamentals (and still with the same $W=M q {\tilde q}$ superpotential for the massless fields). We match it with the low-energy theory we found in theory A for  $m>0$, and obtain the following duality :
\begin{equation}
\begin{aligned}
SU(N)_{k+\frac \fm 2}\quad\longleftrightarrow\quad SU(\nn)_{-k-\frac \fm 2}\times U(1)_{F-N-\frac \fm 2}.
\end{aligned}
\end{equation}

For $m<0$ the ansatz we made does not give vacua that are dual to the one we want, so we need to take a more complicated ansatz. We guess that interesting solutions may be found if some of the eigenvalues of $\phi$ are such that some of the components of the quarks on which $SU(F_m)$ acts remain massless. So the next simplest ansatz we try is of the form
\begin{align} \label{two block}
\phi & =m\begin{pmatrix}
  & -\frac {\fm}{2\left(F-N \right)}\id_{C\times C}\\
  &  & -\left[\frac {\fm}{2\left(F-N \right)}-1 \right]\id_{P\times P}
\end{pmatrix},
\end{align}
with $C+P={\tilde N}$. This ansatz breaks $U({\tilde N})$ to $U(C)\times U(P)$. The $SU(C)\times SU(P)$ factors obviously have CS level $(-k)$, but the $U(1)$ factors come from a combination of the original $U(1)\subset U({\tilde N})$ and a generator of $SU({\tilde N})$, so we need to compute their CS levels by taking the appropriate combinations. We find that the $U(1)\subset U(C)$ has a CS level $(C-k)$, the one in $U(P)$ has a CS level $(P-k)$, and there is also a mixed CS term mixing the two $U(1)$'s, of level one. Including the CS contributions and the appropriate induced FI terms in the vacuum \eqref{two block}, this leads to the following D-term equations for the two $U(1)$'s :
\begin{align}  \label{U(C) equation}
U(1) \subset U(C): 
 &-\left(C-k \right)\frac {m\,\fm}{4\pi\left(F-N \right)}-\frac{mP}{2\pi}\left(\frac \fm{2\left(F-N \right)}-1 \right) \nonumber\\
&\qquad +\frac{\fm\,m\,\sign m}{4\pi}
- \sum_{c=1}^C \left[\sum_{f=F_m+1}^{F} q^{c}_{f}q^{*f}_{c}- \sum_{f=1}^F \tilde q_{c}^{f}\tilde q_{f}^{*c} \right]=0,\\
\nonumber\label{U(P) equation} U(1) \subset U(P): & 
 -\left(P-k \right)\frac{m}{2\pi}\left(\frac \fm{2\left(F-N \right)}-1 \right)-\frac{mC}{2\pi}\frac \fm{2\left(F-N \right)} \\
&\qquad-\frac{\fm\,m\,\sign m}{4\pi}
 -\sum_{c=C+1}^{\tilde N} \sum_{f=1}^{F_m} q^{c}_{f}q^{*f}_{c}=0,
\end{align}
where we included vacuum expectation values (VEVs) just for the fields that remain massless in the vacuum \eqref{two block}. Note that for $P=0$
\eqref{U(C) equation} reduces to the previous case.

Now, we find that there is no solution with vanishing expectation values for the quarks charged under $SU(F-F_m)\times SU(F)_R$ when $m>0$, but there is a solution for $m<0$ and $P=\fm$, if $\fm\le k$ and $\fm \leq {\tilde N}$. In this vacuum we get at low energies a $U(\nn-\fm)$ gauge theory with a shifted CS level $\Delta k=\frac \fm 2$ and with $F-\fm,F$ flavors; the $U(\fm)$ sector is broken due to non-zero VEVs for the quarks. Matching the low-energy theory with theory A now gives :
\begin{equation}
\begin{aligned}
SU(N)_{k-\frac \fm 2}\quad\longleftrightarrow\quad SU(\nn-\fm)_{-k+\frac \fm 2}\times U(1)_{F-N-\frac \fm 2}.\label{fm negative m}
\end{aligned}
\end{equation}

In terms of the low-energy theories, the two cases we discussed until now give precisely the same duality
\begin{equation}
\begin{aligned}
SU(N)_{\tilde k}\quad\longleftrightarrow\quad SU(\nn)_{-\tilde k}\times U(1)_{\nn-\tilde k}
\end{aligned}
\end{equation}
between theories with ${\hat F}$ fundamental and ${\hat {\tilde F}}$ anti-fundamental flavors, where $\tilde k$ is the CS level of the low-energy theory, we have ${\hat F} \times {\hat {\tilde F}}$ mesons on the right-hand side coupled by a $W = M q {\tilde q}$ superpotential, $\nn=\frac{{\hat F}+{\hat {\tilde F}}}{2}+\tilde k-N$, and where in both derivations $\fm=|{\hat F}-{\hat {\tilde F}}|$ obeys $\fm\le 2\tilde k$.
As usual we expect that this duality makes sense, and our classical solutions lead to quantum vacua, when $\nn \geq 0$, while otherwise the theory on the left-hand side breaks supersymmetry.

We still did not find any solution in theory B when $m < 0$ and $\fm > k$, so in this case we need to take a more complicated ansatz. We assume that $\phi$ has an additional block with some eigenvalue $(-m\chi)$, namely that
\begin{align}
\phi & =m\begin{pmatrix}
 -\frac {\fm}{2\left(F-N \right)}\id_{C\times C}\\
  & -\left[\frac {\fm}{2\left(F-N \right)}-1 \right]\id_{P\times P}\\
& &-\chi \id_{\nn-C-P}\\
\end{pmatrix}.
\end{align}
The gauge group is now broken as
\begin{equation}
\begin{aligned}
U(\nn)\rightarrow U(C)\times U(P)\times U(\nn-C-P).
\end{aligned}
\end{equation}

As before, we find that the CS level for the $U(1)$ factor in each block of size $L$ is $(L-k)$, and that there is a mixed CS coupling for the $U(1)$'s in each pair of blocks. The $U(1)$ D-term equations, taking into account the CS terms and the induced FI terms, are thus now:
\begin{align}\label{U(C) equationn}
& U(1) \subset U(C): \nonumber
 -\left(C-k \right)\frac {m\,\fm}{4\pi\left(F-N \right)}-\frac{mP}{2\pi}\left(\frac \fm{2\left(F-N \right)}-1 \right)-\frac{m\chi}{2\pi}\left(\nn-C-P \right)\\
&\qquad\qquad\qquad\quad+\frac{\fm\,m\,\sign m}{4\pi}-\sum_{c=1}^{C}\left[\sum_{f=\fm+1}^{F}q^{c}_{f}q^{*f}_{c}-\sum_{f=1}^{F}\tilde q^{f}_{c}\tilde q^{*c}_{f}\right]=0,\\
& U(1) \subset U(P): \label{U(p) equationn}\nonumber
 -\left(P-k \right)\frac{m}{2\pi}\left(\frac \fm{2\left(F-N \right)}-1 \right)-\frac{m\chi}{2\pi}\left(\nn-C-P \right)-\frac{mC}{2\pi}\frac \fm{2\left(F-N \right)}\\
 &\qquad\qquad\qquad\quad-\frac{\fm\,m\,\sign m}{4\pi}-\sum_{c=C+1}^{P}\sum_{f=1}^{\fm}q^{c}_{f}q^{*f}_{c}=0,\\
& U(1) \subset U(\nn-C-P): \label{U(nn-c-p) equation}\nonumber
 -\left(\nn-C-P-k \right)\frac{m\chi}{2\pi}-\frac{m\,\fm\,C}{4\pi\left(F-N \right)}\\
&\qquad\qquad\qquad\qquad\qquad-\frac{mP}{2\pi}\left(\frac \fm{2\left(F-N \right)}-1 \right)+\frac{\fm\,m\,\sign m}{4\pi}\cdot G=0,
\end{align}
where $G$ is given by:
 \begin{equation}
 \begin{aligned}\label{Big G}
 G=\left(\frac \fm{2\left(F-N \right)}-\chi-1 \right)\sign{\frac \fm{2\left(F-N \right)}-\chi-1}\\
 \qquad -\left(\frac \fm{2\left(F-N \right)}-\chi \right)\,\sign{\frac \fm{2\left(F-N \right)}-\chi} .
 \end{aligned}
 \end{equation}

In order to get the desired flavor symmetries, we are looking for a solution of the form $q=\tilde q=0$ in \eqref{U(C) equationn}. We find that when the third block is present,
the only consistent case is precisely the missing case $\fm>k,\,m<0$ and then there is a solution when $P=0$ and $\chi<\frac{\fm}{2\left(F-N \right)}-1$.
The equations then give $C=F-N$, so the $\chi$ block is of size $k$. 

At low energies we get in this block a $SU(k)_{-k}\times U(1)_0$ theory with no charged matter fields, where the $SU(k)_{-k}$ part is at low energies a trivial decoupled theory. The $U(1)$ part has a mixed CS level with the $U(F-N)$ sector. Thus, we get at low energies a $SU(F-N)_{-k+\frac\fm 2}\times U(1)_{F-N-k+\frac \fm 2}\times U(1)_0$ theory, where the matter is charged under the first two factors, and the third is coupled to the second by a mixed CS coupling. 

By considering the low-energy theories, the duality for this case can be written as (still with mesons on the right-hand side coupled by a $W = M q {\tilde q}$ superpotential)~:
\begin{equation}
\begin{aligned}
SU(N)_{\tilde k}\quad\longleftrightarrow\quad SU(F_{\te {max}}-N)_{-\tilde k}\times U(1)_{F_{\te {max}}-N-\tilde k}\times U(1)_0,
\end{aligned}
\end{equation}
with ${\hat F}, {\hat {\tilde F}}$ flavors, 
where $F_{\te {max}}=\max({\hat F},\hat {\tilde F})$ and $\fm=|{\hat F} - {\hat {\tilde F}}|>2\tilde k$, where $\tilde k$ is the CS level of the low energy theory.

As in theory A, there are also various other supersymmetric vacua, which generally give purely topological theories at low energies. We expect these to exactly match with the extra vacua of theory A, but we did not explicitly verify this.

To summarize, our main result is that for an $SU(N)_k$ Chern-Simons gauge theory with $k>0$ and $F,\tilde F$ fundamental and anti-fundamental flavors, the dual theory is:
\begin{equation}
\begin{aligned}\label{new duality}
&SU(\nn)_{-k}\times U(1)_{\nn-k},&\qquad \Delta F\le 2k\\
&SU(F_{\te {max}}-N)_{-k}\times U(1)_{F_{\te {max}}-N-k}\times U(1)_0,&\qquad \Delta F> 2k
\end{aligned}
\end{equation}
with $F,\tilde F$ fundamental and anti-fundamental flavors, where $\Delta F=|F-\tilde F|$, we have $F\times {\tilde F}$ mesons coupled by a $W = M q {\tilde q}$ superpotential, in the first line the rank $\nn=\frac{F+\tilde F}2+k-N$, and in the second line $F_{\te {max}}=\max(F,\tilde F)$ and there is a mixed CS term coupling the two $U(1)$'s of level one. For $\tilde F=0$ this gives a different dual than the previously known one \cite{Aharony:2013dha}, and we expect the relation between the two to be similar to other cases with two different duals discussed in \cite{Aharony:2013dha}.

At the end of the day, the dualities we find are very similar to the $U(N)$ dualities of \cite{Benini:2011mf}. This suggests that it may be possible to derive them by `ungauging' the $U(1)$ in the $U(N)$ dualities, along the lines discussed in \cite{Aharony:2013dha,Park:2013wta}. However, because of the Chern-Simons term for the $U(1) \subset U(N)$, it is not obvious exactly how to do this.
Note that in the $U(N)$ duality of \cite{Benini:2011mf} there are extra terms in the superpotential
involving monopole operators when $|\Delta F| = 2k$, related to the fact that in
that case there is an uncharged monopole. However, for $SU(N)$ this does not happen.

\section{A test of the duality}
\label{flat}

Most tests of the new duality \eqref{new duality} work in a straightforward way. In this section we describe in detail how the baryonic flat directions match, since this is less trivial. 

For simplicity, we discuss only the vacua where the gauge-invariant fields $M=0$, while one component of the baryons ($B=Q^N$ or $\tilde B=\tilde Q^{N}$ in theory A) is turned on. Note that $B$ can be non-zero only for $F\ge N$, while $\tilde B$ can be non-zero only for $\tilde F\ge N$.

Recall that for $F=\tilde F$ the baryonic flat directions were matched in \cite{Aharony:2013dha}. In theory B they involve giving a VEV of rank $k$ to the vector multiplet, such that $B^k$ maps to the monopole operator
\begin{equation}
\begin{aligned}
B^k\simeq q^{(F-N)k} \exp\left(\frac{\phi_1+\cdots+ \phi_k}{g^2}+i\left( a_1+\cdots+ a_k \right) \right).
\end{aligned}
\end{equation}
Here the $\phi_i$'s are the eigenvalues of $\phi$, and the $a_i$'s are the duals to the photons when the gauge group is broken to its Cartan subgroup. We want to study how this mapping works for the general $F\ne \tilde F$ case.

In theory A, it is obvious that the $SU(N)$ D-term equations (when we do not turn on a VEV for the vector multiplet)
\begin{equation}
\begin{aligned}
\alpha\delta^{c}_{c'} & =\sum_{f=1}^FQ^{*c}_{f}Q^{f}_{c'} - \sum_{f=1}^{\tilde F}\tilde{Q}_{c'}^{f}\tilde{Q}_{f}^{*c},
\end{aligned}
\end{equation}
for $c,c' = 1,\cdots,N$, have a solution with non-zero $B$ when $F \ge N$, and with non-zero ${\tilde B}$ when ${\tilde F} \ge N$.
Note that on the baryonic branch the gauge group is broken at a high scale if the VEVs are large, so the classical analysis of the vacua is valid.

The form of the corresponding solutions in theory B depend on the relative value of $\Delta F\equiv |F-\tilde F|$ and of the CS level $k$. Without loss of generality we assume $F<\tilde F$ (the other case is related to this by charge conjugation).

We start with the $|\Delta F|\le 2k$ case. Theory B is given on the first line of \eqref{new duality}.
Given the form of the matching of the  baryonic branches in the $\Delta F=0$ case, we expect that we should turn on a non-zero block in $\p$. So we use the following minimal ansatz:
\begin{align}
\phi & =
\begin{pmatrix}
	\chi \id_{C}\\
	& 0\cdot \id_{P}
\end{pmatrix},
\end{align}
where $C+P=\nn$. The D-term equations for the two $U(1)$'s are derived by an analysis similar to the one in the previous section, 
\begin{equation}
0=\left(C-k \right)\frac{\chi}{2\pi}-\frac{\Delta F\,\chi\,\sign\chi}{4\pi},
\end{equation}
\begin{equation}
0=\frac{C\chi}{2\pi}-\sum_{c=C+1}^{\tilde N} \left[ \sum_{f=1}^F q^{*f}_{c}q^{c}_{f}- \sum_{f=1}^{\tilde F} \tilde{q}_{f}^{c}\tilde{q}_{c}^{f*} \right].
\end{equation}
It is clear that indeed we must have a non-trivial block ($C > 0$), since otherwise  we need to turn on both $q$ and $\tilde q$ to have a non-trivial vacuum.
From here
we obtain
equations for $C$ and $P$:
\begin{equation}
C=k+\frac {\Delta F} 2\sign \chi, \qquad
P=F-N+\frac {\Delta F} 2-\frac{\Delta F\,\sign \chi}{2}.
\end{equation}
 
If we turn on a VEV for the $q$'s then $\chi>0$ and $P=F-N$; we have a one dimensional branch that satisfies (in order to satisfy the $SU(P)$ and $U(1)$ D-terms) $F\ge F-N\ge 0$ in agreement with theory A, in which the rank of $\phi$ is equal to $k+({\tilde F}-F)/2$. If we turn on a VEV for the $\tilde q$'s then $\chi<0$, $P=\tilde F-N$ and the $SU(P)$ rank inequality $\tilde F \ge \tilde F-N\ge 0$ must hold, again in agreement with theory A. In this case the rank of $\phi$ is $k+(F-{\tilde F})/2$. All in all we find in this case a straightforward generalization of the $\Delta F = 0$ case.

Next we consider the $\Delta F>2k$ duality on the second line of \eqref{new duality}, which gives ${\tilde N} = {\tilde F} - N$. We denote by $\eta$ the VEV of the scalar in the vector multiplet of the last $U(1)$. If $\phi=0$ then the D-term equation of the last $U(1)$ is trivially satisfied, and the first one gives
\begin{equation}
\begin{aligned}
0&=\frac{k\eta}{2\pi}-\sum_{c=1}^{\tilde N} \left[ \sum_{f=1}^F q^{f*}_{c}q^{c}_{f}-\sum_{f=1}^{\tilde F} \tilde{q}_{f}^{c}\tilde{q}_{c}^{f*}\right].
\end{aligned}
\end{equation}
If we want to turn on $\tilde q$ we find that due to the $SU({\tilde N})$ equations, this is possible whenever $\tilde F\ge {\tilde N}=\tilde F-N\ge 0$, as in theory A. However, if we try to turn on a VEV for the $q$'s we find that this is possible only when
$F\ge \tilde F-N \ge 0$, which does not cover all the cases where this branch exists in theory A.

Thus, for the other cases we need to take $\phi \neq 0$, such that the dual to the baryon includes a monopole operator.
The D-term equation of the last $U(1)$ implies ${\rm tr}(\phi)=0$, so the minimal ansatz in this case requires two non-zero blocks:
\begin{align}
\phi & =
\begin{pmatrix}
	\chi \id_{C}\\
	&\tilde\chi \id_{\tilde C}\\
	&& 0\cdot \id_{P}
\end{pmatrix}.
\end{align}
The D-term equations are now
\begin{equation}
\begin{aligned}
0&={C\chi}+{\tilde C\tilde\chi},\\
0&=\left(C-k \right)\frac{\chi}{2\pi}+\frac{\tilde C\tilde\chi}{2\pi}+\frac{k\eta}{2\pi}-\frac{\Delta F\,\chi\,\sign\chi}{4\pi}, \\
0&=\left(\tilde C-k \right)\frac{\tilde\chi}{2\pi}+\frac{ C\chi}{2\pi}+\frac{k\eta}{2\pi}-\frac{\Delta F\,\tilde\chi\,\sign{\tilde\chi}}{4\pi}, \\
0&=\frac{C\chi}{2\pi}+\frac{\tilde C\tilde\chi}{2\pi}+\frac{k\eta}{2\pi}-\sum_{c=C+{\tilde C}+1}^{\tilde N} \left[ \sum_{f=1}^F q^{f*}_{c}q^{c}_{f}- \sum_{f=1}^{\tilde F} \tilde{q}_{f}^{c}\tilde{q}_{c}^{f*} \right].
\end{aligned}
\end{equation}
We get equations for $C$ and $\tilde C$, for which the simplest solution is\footnote{We take without loss of generality $\chi>0$, $\tilde \chi<0$.} :
\begin{equation}
\label{vev for Q}
C=\frac {\Delta F} 2+k, \qquad
\tilde C=\frac {\Delta F} 2-k, \qquad
P=F-N.
\end{equation}
And, we find that we can turn on a VEV for the $q$'s, with $\eta>0$. The rank condition that we find using the $SU(P)$ equations, $F\ge F-N\ge 0$, is consistent with theory A.

Thus, the baryonic flat directions match in all cases. It would be interesting to precisely match also the baryonic operators, but we leave this to future work.

\section*{Acknowledgements}
This work was supported in part by an Israel Science Foundation center for excellence grant (grant no. 1989/14), by the Minerva foundation with funding from the Federal German Ministry for Education and Research, by the I-CORE program of the Planning and Budgeting Committee and the Israel Science Foundation (grant number 1937/12), by a Henri Gutwirth award from the Henri Gutwirth Fund for the Promotion of Research, and by the ISF within the ISF-UGC joint research program framework (grant no. 1200/14). OA is the Samuel Sebba Professorial Chair of Pure and Applied Physics.


\providecommand{\href}[2]{#2}\begingroup\raggedright\endgroup

\end{document}